\documentclass[manuscript]{aastex}
\newcommand{\mdeg}[2]{$#1\mbox{$^\circ \mskip-7.6mu.\,$}#2$} %% \mdeg{3}{31} = grados en formato 3^o .31

\shorttitle{Multi--Epoch VLBA observations of Cyg OB2 \# 5.}
\shortauthors{Dzib et al.}

\begin{document}

\title{Multi--Epoch VLBA Observations of the Compact Wind--Collision Region in
the Quadruple System Cyg OB2 \#5}

\author{Sergio A. Dzib\altaffilmark{1}, Luis F. Rodr{\'{\i}}guez\altaffilmark{1,2},
Laurent Loinard\altaffilmark{1,3},  
Amy J. Mioduszewski\altaffilmark{4},  Gisela N. Ortiz--Le\'on\altaffilmark{1}
and Anabella T. Araudo\altaffilmark{1}}

\email{s.dzib@crya.unam.mx}

\altaffiltext{1}{Centro de Radiostronom\'ia y Astrof\'isica, Universidad Nacional Aut\'onoma de Mexico,
Morelia 58089, Mexico}
\altaffiltext{2}{Astronomy Department, Faculty of Science, King Abdulaziz University,
P.O. Box 80203, Jeddah 21589, Saudi Arabia}
\altaffiltext{3}{Max-Planck-Institut f\"ur Radioastronomie, Auf dem H\"ugel 69, 53121 Bonn, Germany}
\altaffiltext{4}{National Radio Astronomy Observatory, 1003 Lopezville Road, Socorro, NM 87801,
USA}

\begin{abstract}
We present multi--epoch VLBA observations of the compact wind collision
region in the Cyg OB2 \#5 system. These observation confirm the arc-shaped 
morphology of the emission reported earlier. The total flux as a function 
of time is roughly constant when the source is ``on'', but falls below the
detection limit as the wind collision region approaches periastron in its 
orbit around the contact binary at the center of the system. In addition, at
one of the ``on'' epochs, the flux drops
to about a fifth of its average value. We suggest that this apparent variation could
result from the inhomogeneity of the wind that hides part of the flux rather than 
from an intrinsic variation. We measured a
trigonometrical parallax, for the most compact radio emission of 0.61 $\pm$
0.22 mas, corresponding to a distance of 1.65 $^{+0.96}_{-0.44}$ kpc, in
agreement with recent trigonometrical parallaxes measured for objects in the
Cygnus X complex. Using constraints on the total mass of the system and orbital
parameters previously reported in the literature, we obtain two independent 
indirect measurements of the distance to the Cyg OB2 \#5 system, both consistent
with 1.3--1.4 kpc. Finally, we suggest that the companion star responsible for the 
wind interaction, yet undetected, is of spectral type between B0.5 to O8.
\end{abstract}

\keywords{stars: individual (Cyg OB2 \#5) --- stars: winds ---
 radiation mechanisms: nonthermal --- astrometry}

\section{Introduction}

Non-thermal radio emission from a massive binary system is usually 
produced in the interaction of their strong winds (see De Becker\ 2007 
for a recent review).
An interesting case is the quadruple massive system Cyg OB2 \#5 
(V729 Cyg, BD +40 4220) that is the only multiple system known so far 
to harbor two radio--imaged wind--collision regions (Ortiz--Le\'on et 
al.\ 2011).

Cyg OB2 \#5 is a radio-bright massive multiple system located in the
Cygnus OB2 association, one of the most prominent massive young
clusters known, containing over 100 O stars. The most massive 
component of Cyg OB2 \#5 (i.e.\ the primary) is an eclipsing, contact 
binary consisting of two O-type supergiants with a 6.6-day orbital period. 
We shall refer to this central object as Cyg OB2 \#5 A (containing the
individual stars Cyg OB2 \#5 Aa and Cyg OB2 \#5 Ab) . Contreras et 
al.\ (1997) suggested that a radio source located $0\rlap.{''}8$ to the 
north--east of Cyg OB2 \#5 A is the shock formed  by the interaction 
between the wind from Cyg OB2 \#5 A and that of a B--type star first 
reported by Herbig\ (1967) and located $0\rlap.{''}9$ to the north-east.
We shall refer to this B star as Cyg OB2 \#5 C, and to the
wind--collision region between this star and Cyg OB2 \#5 A as  WCR(A-C). 
Recently, Ortiz--Le\'on 
et al.\ (2011) used the Very Long Base Array (VLBA) telescope to obtain
a high angular resolution image ($\sim$10 mas) and reported the
detection of a compact wind--collision region with 
possible non-thermal radio emission. This structure is formed by the 
interaction between the wind driven by Cyg OB2 \#5 A and that of 
an unseen companion (that we shall call Cyg OB2 \#5 B), 
an early B type star, with an orbital period of $\sim6.7$-yr (Kennedy et 
al.\ 2010) \footnote{Kennedy et al.\ (2010) refer to Cyg OB2 \#5 Aa, Ab, B and C 
as Star A, B, C and D, respectively.}. 
As this star orbits around Cyg OB2 \#5 A, the resulting WCR
(hereafter WCR(A-B)) is undetected around what 
is presumed to be orbit periastron, yet can be detected as the orbit progresses 
to apastron. Ortiz--Le\'on et 
al.\ (2011) noticed that the separation between the WCR(A-B) and the 
position of the contact binary is 12 mas, i.e.\ smaller than the nominal 
radius (23 $\pm$ 12 mas; Rodr\'{\i}guez et al.\ 2010) of the region where 
the optically--thick free--free radiation of the Cyg OB2 \#5 A binary system is produced. 
Therefore, the WCR(A-B) emission should in principle be undetectable due to the 
free--free opacity. To resolve this conundrum, these authors proposed 
that the wind driven by the Cyg OB2 \#5 is very inhomogeneous, as 
suggested for other early massive stars (Blomme et al.\ 2010; Muijres et al.\ 2011).
However, the time variability could be due to other plasma effects 
(Pittard \& Dougherty 2006; Pittard 2009).

Being part of the Cyg OB2 region, Cyg OB2 \#5 is expected to be located at the same distance. 
Traditionally, the Cyg OB2 region has been believed to be at a distance of 2.1 kpc (Reddish 
et al. 1966), although it has, more recently, been argued to be somewhat nearer (at about 1.7 kpc;
e.g. Massey \& Thompson 1991). Even shorter distances have been proposed. For instance, 
Hanson (2003) suggested d $\sim$ 1.4 kpc or even less. Direct trigonometric parallaxes were
obtained recently by Rygl et al.\ (2012) who measured the distances to several methanol and water
maser related to objects in the Cygnus X star forming complex. They found that
most of them are consistent with a distance of 1.40 $\pm$ 0.08 kpc, and suggest
that the Cyg OB2 region is located at a similar distance. Using similar techniques, 
Zhang et al.\ (2012) measured the distance to the red hypergiant NML Cyg to be 
1.61 $\pm$ 0.12 kpc and suggested that it lies on the far side of the Cyg OB2 region. Thus, 
most recent, reliable measurements suggest a distance of 1.4 kpc for the Cyg OB2
region as a whole, and for Cyg OB2 \#5 in particular. Interestingly, however, Linder 
et al.\ (2009) obtained a distance estimate for Cyg OB2 \#5 itself of  925 $\pm$ 25 pc 
from a light curve analysis. A direct astrometric study to Cyg OB2 \#5 could help solve the 
discrepancy between the distance suggested by Linder et al.\ (2009) and the recently 
measured distances for other objects in the region.

In this paper, we present the analysis of a series of 12 VLBA observations
of the WCR(A-B) in Cyg OB2 \#5, covering a total time span of 1.6 years. These
data will be used to study the nature of the emission mechanism in the WCR(A-B), and the 
distance to Cyg OB2 \#5. In the past, our team has successfully used 
multi--epoch VLBA observations to measure distances to low and intermediate mass
stars with compact nonthermal radio emission (see Dzib et al.\ 2010 and 2011
for recent results). In the case of Cyg OB2 \#5 the wind
collision region is resolved with VLBA observations, but should still be sufficiently 
compact to enable usable astrometry. A description of the observations, their calibration
and imaging are described in Section 2. The results are presented in Section 3, where we 
also describe the structure and variability of the emission. Additionally a rough estimation of the
trigonometric parallax is also presented in this section. The discussion of the
results is presented in Section 4, and we finish with the conclusions in Section 5.

%%%%%%%%%%%%%%%%%%%%%%%%%%%%%%%%%%%%%%%%%%%%%%%%%%%%%%%%%%%%%%%%%%%%%%%%%%%%%%%%%%%%%%%%%
\section{Observation and data calibration}
\begin{figure*}[!t]
  \centerline{\includegraphics[height=0.7\textwidth,angle=0]{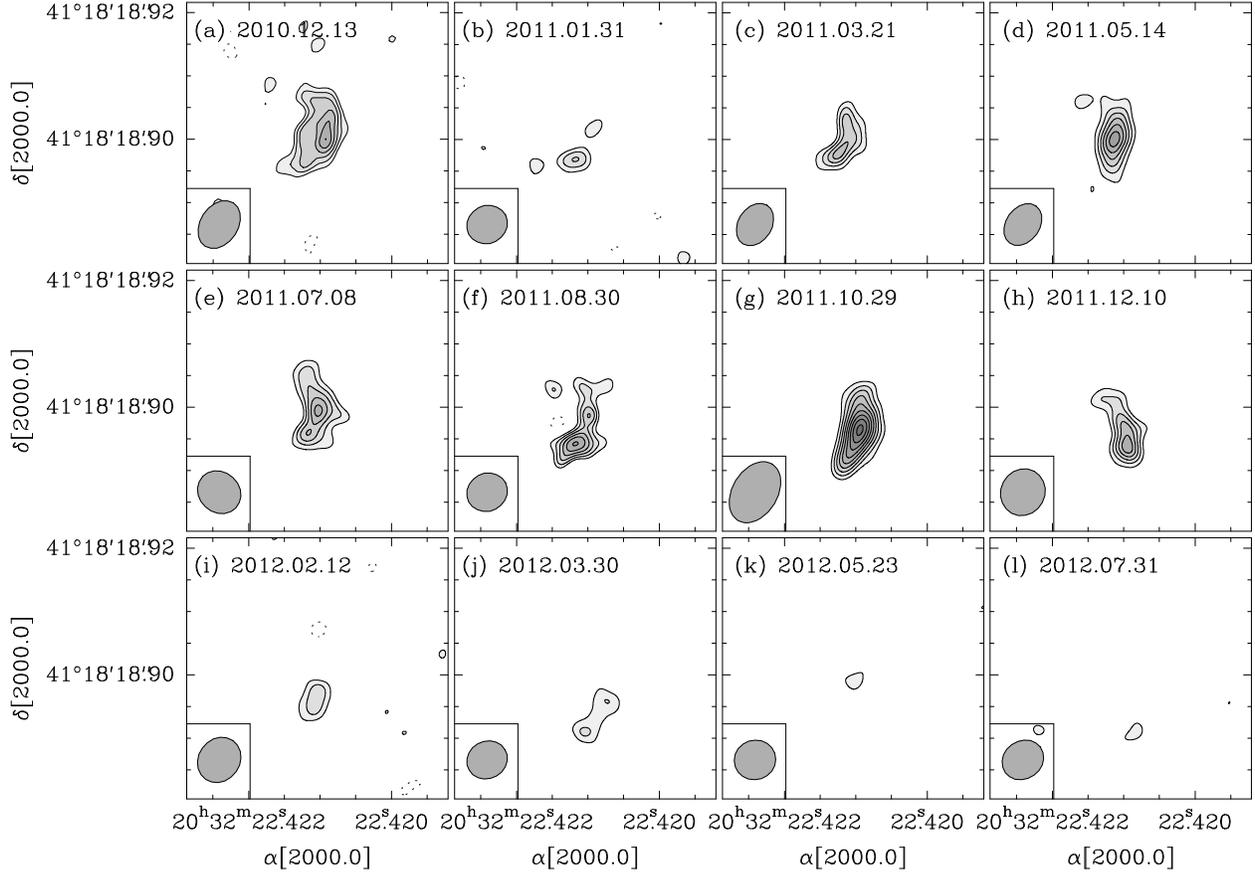}}
  \caption{Images of Cyg OB2 \#5 obtained with the VLBA.
  The contours are at --3, 3, 4.5, 6, 7.5, 9 etc. times the 3$\sigma$ noise
  level in the image (see Table \ref{tab:lri}). The synthesized beam is shown
  at the bottom left.}
    \label{fig:lri}
\end{figure*}

Cyg OB2 \#5 was observed at $\lambda$ = 3.6 cm 
($\nu$ = 8.42 GHz) with the VLBA at 12 epochs between 2010 December and
2012 July. The first observation (in 2010 December) was designed as a detection
experiment and was reported by Ortiz--Le\'on et al.\ (2011). Following this
successful detection, we initiated a series of multi-epoch observations. 
The separation between successive observations in those multi-epoch data was
about 1.5 months (Table \ref{tab:lri}). They were usually realized at a 
recording rate of 512 Mbps
but the epoch 2011 October was recorded using a rate of 2 Gbps, to test 
new VLBA equipment.
As a consequence, the noise level for this epoch was significantly better.

The observations consisted of series of cycles with two minutes spent on source,
and one minute spent on the main phase-referencing quasar J0218+3851, located 
\mdeg{3}{6} away. Every 15 minutes, the radio bright X-ray binary Cyg X-3, located
at $20'$ from Cyg OB2 \#5, was also observed.  Although Cyg X--3 is very close to
Cyg OB2 \#5 it was not used as a primary calibrator due to its extreme variability
in both flux density and morphology and the fact it is heavily scattered at the
observing frequency (Desai \& Fey 2001). As we shall see, however, it was used 
for secondary calibration. The total duration of the observations
was 5 hours for the 2010 December observation and 2 hours for the other epochs.

\begin{figure*}[!t]
  \centerline{\includegraphics[height=0.9\textwidth,angle=-90]{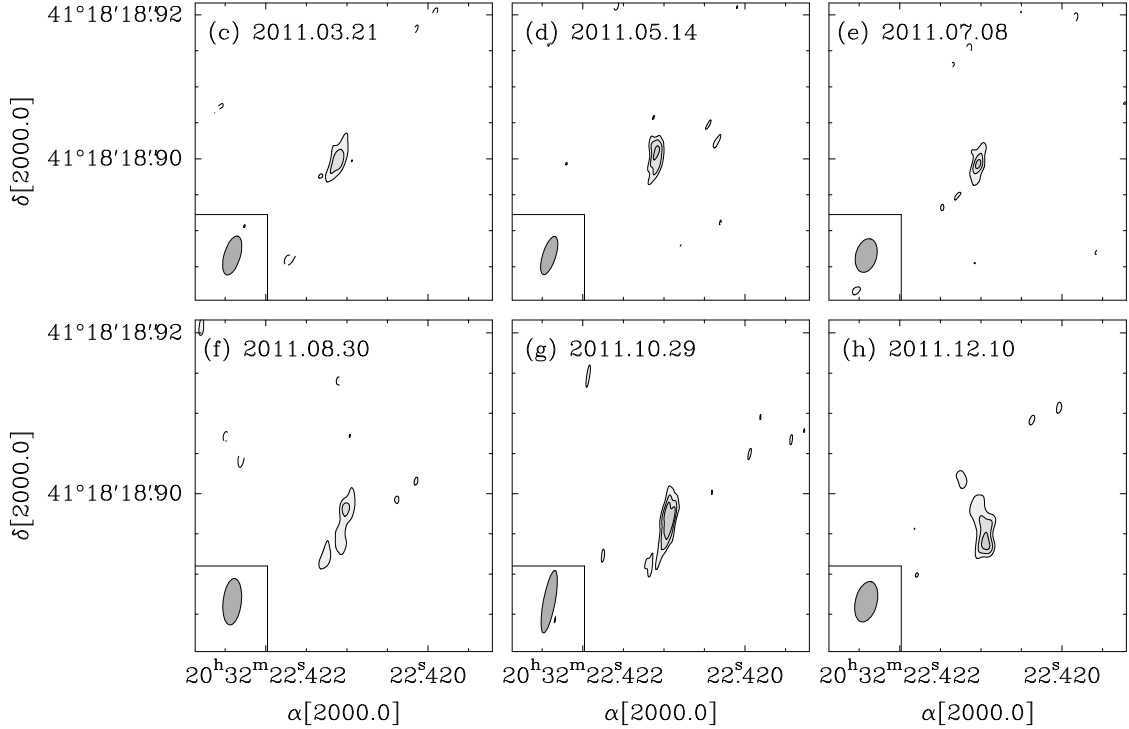}}
  \caption{Images at higher resolution for the epochs when Cyg OB2 \#5 is 
  still detected. The contours are at --3, 3, 4.5, 6, 7.5, 9 etc. times the 3$\sigma$ noise
  level in the image (see Table \ref{tab:hri}). The synthesized beam is shown
  at the bottom left.}
\label{fig:hri}
\end{figure*}

The data were edited and calibrated using the AIPS software, 
following standard procedures for phase--referenced VLBA observations. The 
calibration determined from the observations of J0218+3851 was applied to both 
Cyg OB2 \#5 and Cyg X-3.  At this point, most of the phase errors left are caused 
by the \mdeg{3}{6} separation between the source and the phase calibrator.  To 
remove most of this, Cyg X--3 was self-calibrated in phase and the incremental phase gains
determined from that self-calibration were interpolated and applied to Cyg 
OB2 \#5.  In most of the epochs, this resulted in a significant improvement in 
the quality of the image of Cyg OB2 \#5  compared to a direct phase transfer between
J0218+3851 and Cyg OB2 \#5. For epochs 2011 August and 2011 December, however,
this step worsened the image, presumably because of the high variability of
Cyg X--3. For those epochs, we use the images obtained without applying this step.
Cyg OB2 \#5 is quite resolved, so the best images were made by limiting 
the maximum uv length to 60000 k$\lambda$ and using always natural weighting. 
The total flux for each detection in these images was determined using the AIPS task IMSTAT.
The rms noise level in the final images (shown in Figure \ref{fig:lri}) was 43--110
 $\mu$Jy beam$^{-1}$  (Table \ref{tab:lri}). Additionally, high resolution images 
 were made using the whole uv range. The epochs at which the source was 
 significantly detected in this second set of images are shown in 
Figure \ref{fig:hri}.

As we will see, significant variations in the flux of Cyg OB2\#5 were found. It is,
therefore, worthwhile to examine in some detail the accuracy of the absolute flux
calibration. We followed the standard VLBA recipes, based on gain and system
temperature, for the absolute flux calibration. These methods are applied 
similarly to the target (Cyg OB2 \#5) and the gain calibrator (J0218$+$3851) so
any systematic error should affect equally both sources. We inspected the 
obtained flux of J0218$+$3851 and found no systematic differences from epoch to
epoch. Instead, we found a random scatter of about 0.035 Jy around a mean
value of 0.308 Jy. Thus, the dispersion is about 10\% of the mean and we will 
use that figure as our flux uncertainty.\footnote{Note that J2018$+$3851 is likely 
to be intrinsically moderately variable, and part of the observed scatter might 
reflect this intrinsic variability. As a consequence, our accuracy might well be 
better than the quoted 10\%.}

Given the \mdeg{3}{6} separation between the gain calibrator and the target,
part of the target flux might be lost under adverse weather conditions due to 
decorrelation. This can also be discarded using the observations of Cyg X--3 that 
were intertwined with the observations of Cyg OB2 \#5. Cyg X--3 and Cyg OB2 \#5 
are very near each other (separation of about \mdeg{0}{3}) while the phase 
calibrator is \mdeg{3}{6} degrees away. Thus, if the phase transfer from the 
phase calibrator to the target were responsible for significant decorrelation 
on Cyg OB2\#5, it should also induce significant decorrelation on Cyg X--3. 
We checked that this was not the case in any of our images, by comparing 
the flux measured in our VLBA observations with (publicly available) monitoring 
observations of  Cyg X--3 made with the AMI-large array telescope at 15 GHz.\footnote{
{ The AMI-large array telescope is supported by STFC and the University of 
Cambridge and their monitoring can be accessed by the URL 
http://www.mrao.cam.ac.uk/telescopes/ami/index.html.}}

\begin{table}[!htbp]
\small
  \begin{center}
  \caption{Observed epochs, flux densities {(with statistic and systematic errors)},
 and noise levels of the low resolution VLBA Images.}
    \begin{tabular}{lcccc}\hline\hline
Mean UT date & Modified    & $f_\nu\pm\sigma_{stat}{\pm\sigma_{syst} }$ &$\sigma$    \\
(yyyy.mm.dd) & Julian Day  & (mJy)  & ($\mu$Jy  beam$^{-1}$)\\
\hline
2010.12.13   &   55543.90  & 3.00$\pm$0.13{$\pm$0.30}  &   55    \\%0.95
2011.01.31   &   55592.85  & 0.66$\pm$0.10{$\pm$0.07}  &   80    \\
2011.03.21   &   55641.72  & 2.98$\pm$0.18{$\pm$0.30}  &   89    \\
2011.05.14   &   55695.57  & 3.51$\pm$0.16{$\pm$0.35}  &   86    \\
2011.07.08   &   55750.42  & 3.24$\pm$0.16{$\pm$0.32}  &   76    \\
2011.08.30   &   55803.28  & 2.57$\pm$0.13{$\pm$0.26}  &   72    \\
2011.10.29   &   55863.11  & 1.66$\pm$0.07{$\pm$0.17}  &   43    \\
2011.12.10   &   55906.00  & 2.58$\pm$0.13{$\pm$0.26}  &   68    \\
2012.02.12   &   55969.82  & 1.33$\pm$0.16{$\pm$0.13}  &  110    \\
2012.03.30   &   56016.69  & 0.98$\pm$0.14{$\pm$0.10}  &   70    \\
2012.05.23   &   56070.55  &          $<$0.24             &   80    \\
2012.07.31   &   56139.36  &          $<$0.25             &   84    \\ %0.58
\hline\hline
%\tabnotetext{b}{}
    \label{tab:lri}
    \end{tabular}
  \end{center}
\end{table}

\section{Results}

\subsection{Structure and properties of the emission}

From Figure \ref{fig:lri} we can see that when the source is detected, the structure
is similar to that reported by Ortiz--Le\'on et al.\ (2011): an arc--like morphology,
with apex located to the west, extended mostly along the north--south direction.
Following Ortiz--Le\'on et al.\ (2011), we attribute this morphology to the interaction
between the strong wind driven by the contact binary Cyg OB2 \#5 A located 
about 12 mas to the west of the WCR(A-B) (see Figure 2 in Ortiz--Le\'on et al.\ 2011), 
and the wind driven by an unseen companion (Cyg OB2 \#5 B) located just east of the 
apex (see again Figure 2 in Ortiz--Le\'on et al.\ 2011). This is also the preferred scenario 
of Kennedy et al.\ (2010) to explain the VLA observations of the system.
In the high resolution images (Figure \ref{fig:hri}), most of the flux
is resolved out, and only the most compact radio emission remains detectable. The implied 
brightness temperature in these cases is in the range of 0.5--1.5 $\times10^{7}$ K (Table 
\ref{tab:hri}).

The existing VLA observations of Cyg OB2 \#5 imply that the average
flux from the WCR(A-B) during high state should be 
$\sim4.0$ mJy at 8.4 GHz (as compared to the total flux of
the stellar winds plus WCR(A-B) of $\sim$9 mJy; Kennedy et al.\ 2010). 
Interestingly, however, the fluxes observed for WCR(A-B) 
in the VLBA observations reported here (Table \ref{tab:lri}) are significantly below the value 
of $\sim4.0$ mJy
($\sim$ 3 mJy on average). This could be taken to indicate that the flux in the current ``on'' 
state is systematically lower than the previous ``on'' states. To check this, we searched the 
archive of the Very Large Array (VLA) for recent observations of the system. We found an 
observation of the project 12B--165 taken in 2010 August in C band and D configuration.
The source was detected at 9.89 $\pm$ 0.38 mJy, consistent with previous detection 
at the same band in the same configuration in previous ``on'' states (see Kennedy et al.\ 
2010). We suggest, instead, that about 25\% of the emission is resolved out by the VLBA.

\begin{figure}[!t]
  \centerline{\includegraphics[height=0.45\textwidth,angle=-90]{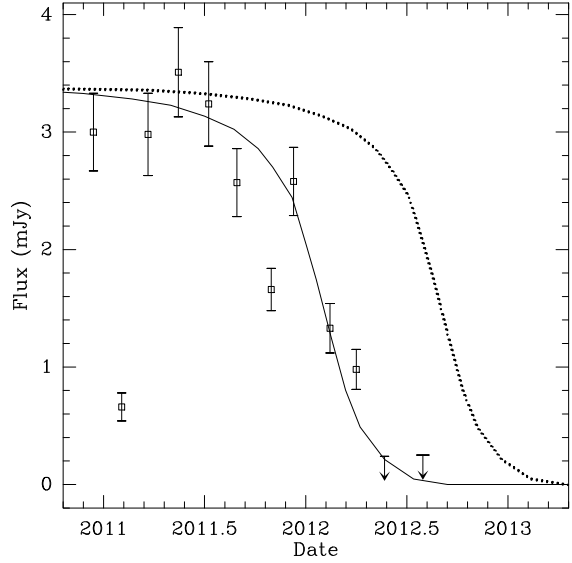}}
  \caption{Time evolution of the total flux of the WCR(A-B) in Cyg OB2\#5, from the low resolution VLBA images.
The upper limits are at 3$\sigma$. The dotted curve shows the expectation from the best fit of Kennedy et al.\
(2010; { the s=0 model}) scaled by 0.75 to account for the missing flux (see text), while the solid curve shows the same 
fit  offset by 7.5 months.} %%%%%%%%%%%%%%%%%%%%%%%%%%%%%%%%%%%%%%%%%%%%%%%%%%%%%%%%
\label{fig:flri}
\end{figure}

In Figure \ref{fig:flri} the fluxes from Table \ref{tab:lri} are plotted as a function 
of time. From December 2010 to December 2011, the flux is roughly constant
at about 3 mJy, except for the observation of 2011 January. After December 2011, 
the flux starts to decrease, and the source is finally not detected in 2012 May and 2012 July. 
This flux behavior is consistent with that expected from previous observations 
of Cyg OB2 \#5, and {\em qualitatively} in agreement with the best formal model (i.e. s=0) of 
Kennedy et al.\ (2010). Quantitatively, however, the best model of Kennedy et al.\ (2010)
predicted the flux to drop about 7.5 months later than it actually did (see Figure \ref{fig:flri}).
%This discrepancy might reflect the fact that the model of Kennedy
%et al.\ (2010) has to be extrapolated to be applied
%to our observations, and additional observations are required
%to better understand it.

\subsection{Astrometry}
For astrometric studies, point-like objects are preferred, and the WCR(A-B) in Cyg 
OB2 \#5 is clearly not the ideal type of target. However, the six high-resolution images 
shown in Figure \ref{fig:hri} can be used to perform some astrometry, albeit with 
less accuracy than would be achievable for very compact sources. In particular, the 
extension of the source in declination induces large positional errors along that 
direction. Although it does include two successive epochs of maximum parallactic 
elongation (March and September), the total time span covered by our observations is 
only 8.8 months, and is also not optimal. Bearing in mind these limitations, we attempted an 
astrometric study of the WCR(A-B) in Cyg OB2 \#5.

\begin{table*}[!htbp]
%\footnotesize
\scriptsize
  \begin{center}
  \caption{Calendar dates and their corresponding Julian days, measured source positions, flux densities, noise levels, and brightness 
temperatures of the high resolution VLBA images.}
    \begin{tabular}{lcccccccc}\hline\hline
Mean UT date &Julian Day & $\alpha$(J2000.0) & $\sigma_{\alpha}$ & $\delta$(J2000.0) & $\sigma_\delta$ & $f_\nu$ &$\sigma$&T$_b$\\
(yyyy.mm.dd/hh:mm)& & $20^{\rm h}32^{{\rm m}}$& & $+41^{\circ}18'$& & (mJy)& ($\mu$Jy beam$^{-1}$)& ($10^7$ K)\\
\hline
2011.03.21/17:17&2455642.22&22\rlap.{$^{\rm s}$}421114 &0\rlap.{$^{\rm s}$}000018&18\rlap.{''}89983&0\rlap.{''}00042&2.17$\pm$0.43&82&1.24\\
2011.05.14/13:44&2455696.07&22\rlap.{$^{\rm s}$}421094 &0\rlap.{$^{\rm s}$}000012&18\rlap.{''}90079&0\rlap.{''}00039&2.57$\pm$0.42&75&1.55\\
2011.07.08/10:08&2455750.92&22\rlap.{$^{\rm s}$}421036 &0\rlap.{$^{\rm s}$}000013&18\rlap.{''}89937&0\rlap.{''}00034&1.32$\pm$0.27&76&0.56\\
2011.08.30/06:40&2455803.78&22\rlap.{$^{\rm s}$}420999 &0\rlap.{$^{\rm s}$}000018&18\rlap.{''}89792&0\rlap.{''}00046&1.37$\pm$0.32&81&1.02\\
2011.10.29/02:44&2455863.61&22\rlap.{$^{\rm s}$}420940 &0\rlap.{$^{\rm s}$}000012&18\rlap.{''}89689&0\rlap.{''}00042&1.40$\pm$0.22&41&0.74\\
2011.12.10/23:55&2455906.50&22\rlap.{$^{\rm s}$}420949 &0\rlap.{$^{\rm s}$}000015&18\rlap.{''}89466&0\rlap.{''}00050&2.19$\pm$0.44&68&0.89\\
\hline\hline
    \label{tab:hri}
    \end{tabular}
  \end{center}
\end{table*}

\begin{figure}[!t]
  \centerline{\includegraphics[height=0.5\textwidth,angle=-90]{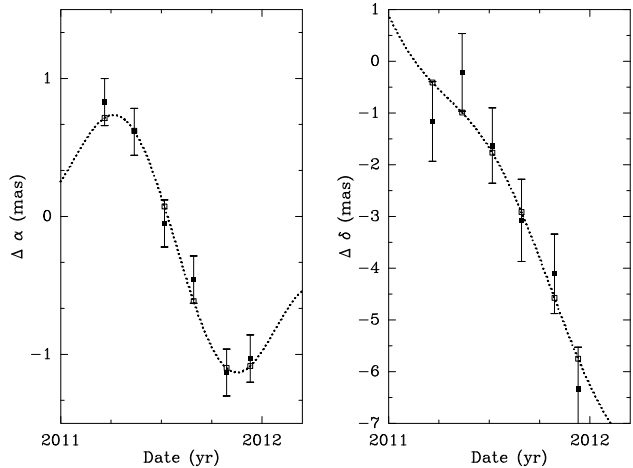}}
  \caption{Best astrometric fit to the positions obtained from the high 
resolution VLBA images. The left panel shows the right ascension as a function
of time, while the right panel shows the declination behaviour. The filled
squares show the observed positions with their error bars (including the 
systematic contributions) while the open squares indicate the position of the source
expected from the fits at each epoch. The dotted curves are the best fits to the data
including a parallax and uniform proper motion.}
\label{fig:parx}
\end{figure}

The positions of the source in our VLBA data were determined 
from the high resolution images shown in Figure \ref{fig:hri} using a 
two--dimensional Gaussian fitting procedure (task JMFIT in AIPS) 
and are given in Table \ref{tab:hri}. JMFIT provides an estimate of 
the position error based on the expected theoretical astrometric 
precision of an interferometer (Condon 1997); these errors are 
quoted in Columns 4 and 6 of Table \ref{tab:hri}. To obtain the 
astrometric parameters from these data, we used the
single value decomposition fitting scheme described by Loinard
et al. (2007). The necessary barycentric coordinates of the Earth,
as well as the Julian date of each observation, were calculated
using the Multi-year Interactive Computer Almanac (MICA)
distributed as a CD ROM by the US Naval Observatory. The
reference epoch was taken at the mean of the four observations: JD
2455774 $\equiv$ 2011.6. The best fit is shown
in Figure \ref{fig:parx}, and yields the following astrometric elements:

\begin{eqnarray*}
\alpha_{J2011.6} & = & 20^{{\rm h}}32^{{\rm m}}22\rlap.{^{\rm s}}421021\pm 0.000006\\
\delta_{J2011.6} & = & 41^{\circ}18^{'}18\rlap.{''}89823 \pm 0.00032\\
\mu_\alpha \cos{\delta}&=&-1.64 \pm 0.98\ {\rm mas\ yr}^{-1}\\
\mu_\delta&=&-7.16 \pm 1.28\ {\rm mas\ yr}^{-1}\\
\pi&=&0.61 \pm0.22\ {\rm mas,}
\end{eqnarray*}

\noindent
corresponding to a distance of 1.65$_{-0.44}^{+0.96}$ kpc. This is in agreement
with the recent distances measured for sources related with the Cygnus--X
complex (Zhang et al.\ 2012; Rygl et al.\ 2012) and significantly larger
than the value obtained by Linder et al.\ (2009) for Cyg OB2 \#5 itself. The
post-fit rms in this case is 0.13 mas in right ascension and 0.55 mas in
declination. The reduced-$\chi^2$ obtained in right ascension using the errors
delivered by JMFIT is 0.59, suggesting that no significant systematic
errors remain in our data along that axis. In declination, however, a systematic
contribution of 0.65 mas has to be added quadratically to the errors given
by JMFIT to obtain a reduced-$\chi^2$ of 1. This is not surprising given the
elongation of the source in the north-south direction. All the errors quoted 
here include this systematic contribution.

\section{Discussion}

\subsection{Nature of the radio emission from the WCR(A-B)}

The emission from the WCR(A-B) has long been thought to be of non-thermal origin
because (i) the flat radio spectrum of Cyg OB \#5 in VLA observations when this
source is ``on'' compared to its positive spectral index when it is ``off'' suggests 
that the spectral index of the WCR(A-B) is negative, and (ii) Ortiz--Le\'on et al.\ 
(2011) measured a brightness temperature of $\sim2\times10^6$ K suggestive 
of a non-thermal process. The brightness temperatures reported here are about 
$10^7$ K, clearly favoring a non-thermal (presumably synchrotron) nature.

\begin{figure}[!t]
  \includegraphics[trim = 0mm 5mm 0mm 0mm, clip,height=0.45\textwidth,angle=0]{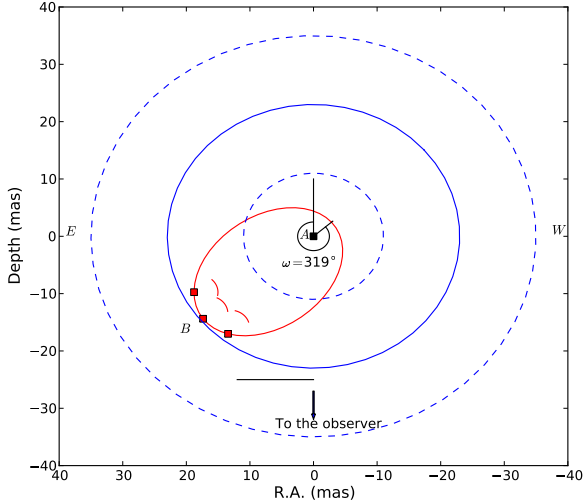}
  \caption{Graphical sketch of the orbit Cyg OB2 \#5 B (red line).
 The blue circle shows the extent of the optically--thick photo--ionized region surrounding Cyg OB2 \#5 A (the full line 
 is the nominal value, while the dashed versions show the minimum and maximum values; 
 from Rodr\'{\i}guez et al.\ 2010)).
 As the sizes depends on the distance assumed, the values of both R.A. and depth are presented here
 in angular units (mas). The black line is the angular separation 
 observed in the plane of the sky. The red squares shows the position of Cyg OB2 \#5 B
 at three representative epochs when the WCR(A-B) can be detected to show the direction
 of the curvature of the WCR(A-B) (red arcs) in these epochs.} 
%%%%%%%%%%%%%%%%%%%%%%%%%%%%%%%%%%%%%%%%%%%%%%%%%%%%%%%%%%%%%%%%%%%%%%%%%%%%%%%%%%%%%%
\label{fig:orb}
\end{figure}

{ The best-fit model of Kennedy et al.\ (2010) has orbital parameters (orbit 
inclination $i=90^\circ\pm40^{\circ}$, argument of periastron 
$\omega=319^{\circ}\pm3^{\circ}$ and eccentricity $e=0.7\pm0.04$) that }result in
a structure as shown in Figure  \ref{fig:orb} (the plot is made in the plane of
the orbit, perpendicular to the plane of the sky). The large circle shows the 
nominal size (radius 23 mas) of the free-free emission region from the Cyg OB2 
\#5 A contact binary together with its uncertainty (12 mas; Rodr\'{\i}guez et al.\ 
2010). It is clear that near apastron (and only near apastron), Cyg OB2 \#5 B (and 
its associated WCR(A-B)) are located outside, or at least near the outer edge, of the 
optically--thick ionized region. This occurs because the orbit is quite eccentric, and explain why 
the non-thermal emission is detectable in spite of the small separation between 
the WCR(A-B) and the contact binary. Note that since Cyg OB2 \#5 B will always be 
located {\bf to the east} of Cyg OB2 \#5 A at apastron, the curvature of the WCR(A-B) 
is expected to always be oriented in nearly exactly the same direction, as observed 
in our VLBA data. On the other hand, we note that the decline in flux of the WCR(A-B)
around periastron can be influenced by other plasma and radiation processes 
(Pittard \& Dougherty 2006; Pittard 2009).

As we mentioned before, the flux of the WCR(A-B) is consistent with a value
of 3 mJy from 2010 December to 2011 December. There is, however, one
exception: the flux in 2011 January is nearly 5 times smaller than this
average value. Our discussion of the absolute flux accuracy (end of 
Section 2) shows that our fluxes are accurate to about 10\% so the
decrease by a factor of 5 observed at that epoch cannot be ascribed to 
instrumental effects. Such strong variability is not expected in a wind collision
region. { The s=0 model of Kennedy et al.\ (2010) predicts that
the flux of the WCR(A-B) is closely constant in the "on" state.} A possible
solution to this discrepancy would be for the wind driven by the contact
binary to be inhomogeneous. Under this assumption, two scenarios would be
possible. The first one would be for Cyg OB2 \#5 B to pass through a region
where the wind driven by Cyg OB2 \#5 A is less dense. In such a situation,
the shock resulting from the wind collision might be weak, and insufficient
to produce relativistic electrons and, therefore, detectable compact
radio emission. Araudo et al.\ (2012, in preparation) computed the cooling
time for the relativistic electrons at the WCR(A-B) considering synchrotron
losses, Inverse Compton (IC) scattering, relativistic bremsstrahlung, and
advection and diffusion escape of the relativistic particles from the
shocked winds.
They found that due to the dense photon stellar field provided by the  
unseen companion, IC  scattering is the most important cooling channel for the
relativistic electrons at energies above $\sim 10^6$ eV
(considering different spectral types for Cyg OB2 \#5 B), 
implying that most of the relativistic electrons are cooled by IC losses. 
Below this energy,
the cooling is dominated by advection of the particles on a time $t_{\rm esc}
\sim 4 l_{\rm bs}/v_{\infty} \sim 4$ months. Here $l_{\rm bs}$ is the size
of the WCR(A-B) and $v_{\infty}$ the terminal
velocity of the stellar winds. Since the separation between our successive 
observations is significantly shorter than this, we would not expect the
flux to drop appreciably between observations.
 A second way to produce a drop in radio flux would be to have
a sudden increase in the column density of ionized material along the line of
sight to the WCR(A-B), for instance as a result of a dense clump passing in front
of the WCR(A-B). This would result in an increase in the free-free opacity. This
second scenario fits more naturally the observed light curve.

\subsection{The distance to Cyg OB2 \#5}

As we mentioned in the Introduction, there is some significant uncertainty on the 
distance to Cyg OB2 \#5. On the one hand, recent  trigonometric parallax
measurements to masers associated with several stars in the Cygnus X 
star forming complex, including the Cygnus OB2 star forming region, suggest 
a distance of 1.4--1.6 kpc (Rygl et al.\ 2012; Zhang et al.\ 2012). On the other
hand, Linder et al.\ (2009) obtained a distance modulus for Cygnus OB2 \#5 
that yields a significantly shorter distance of 925 pc. The latter authors noted 
that theirs results were strongly dependent on the assumed stellar effective 
temperature. The value used was 36,000 K, as obtained from a spectrum 
analysis carried by Rauw et al.\ (1999), that suggested an O6--7 Ia star. They 
also mention that a somewhat unlikely effective temperature of 48000 K would 
be needed if the system were located to a distance of 1.7 kpc. The parallax 
obtained here, although uncertain, favors a distance of order 1.6 kpc, in 
agreement with those obtained by Rygl et al.\ (2012) and Zhang et al.\ 
(2012) and significantly larger than the distance of 925 pc proposed by 
Linder et al.\ (2009)

The distance to the Cyg OB2 \#5 system can also be estimated using an orbital
parallax method. From our VLBA observations and the argument of periastron
($\omega$) and inclination ($i$) of the Cyg OB2 \#5 AB orbits derived by Kennedy 
et al.\ (2010), we can derive 
the de-projected angular separation between Cyg OB2 \# 5 A and Cyg OB2 \# 5 
B. In all four models discussed by
Kennedy et al. (2010) the inclination is close to 90$^\circ$ (in the
range of 85$^\circ$ to 90$^\circ$)
and can be ignored in the derivation. Then, the de-projected angular distance
is simply equal to the { projected angular distance divided by sin(360$^\circ$-$\omega$)}.
For the case of the best fit model s=0, we find that 
at apastron the de-projected
angular distance is 18.3 mas\footnote{Strictly, this is the separation between Cyg OB2 \# 5 
A  and the WCR(A-B) associated with Cyg OB2 \# 5 B, { but Ortiz--Le\'on 
et al.\ (2011) argue that Cyg OB2 \#5B is very near the apex of the WCR(A-B).}}. 
The total mass of Cyg OB2 \#5 A was calculated to be 41.5 $\pm$ 3.4 M$_\odot$ by 
Linder et al.\ (2009) and the mass of Cyg OB2 \#5 B was estimated to be 
23$^{+22}_{-14}$ M$_\odot$ by Kennedy et al.\ (2010). Thus, the appropriate mass 
range for the Cyg OB2 \#5 AB system is between 50 and 90 M$_\odot$. From the orbital 
parameters given by  Kennedy et al.\ (2010), we can calculate the true separation (in cm) 
at apastron for these two masses, and from the comparison between the true separation 
and the de-projected angular separation, deduce the distance to the system. We obtain 
distances between 1.26 kpc and 1.53 kpc. This is in excellent agreement with the parallax
determination obtained earlier, and with the distance to Cyg OB2 estimated by
Rygl et al. (2012) and Zhang et al.\ (2012). We note, on the other hand, that if we assume 
a distance of 0.95 kpc, the total mass of the triple system would have to be only 21.4 \
M$_\odot$, lower than the primary alone. At the other extreme, assuming a distance of 
1.7 kpc would yield a dynamical mass of 122.4 M$_\odot$, larger than
usually expected. Unfortunately, this method depends critically on the orbital
parameters, { and adoption of any other models in } Kennedy et al. (2010) will
give a different distance.

Yet another (independent) distance determination can be obtained as follows. The separation 
between Cyg OB2 \#5 B and the WCR(A-B) can be estimated from a momentum ratio analysis. 
From Cant\'o et al.\ (1996), we can relate the separation $R_0$ between Cyg OB2 \#5 B and 
the WCR(A-B) to the separation $R_1$ between the WCR(A-B) and Cyg OB2 \#5 A by:

\begin{equation}
R_0=\eta^{1/2}R_1,
\label{eq:winds}
\end{equation}

\noindent
where $\eta=\dot{M}_{w0}V_{w0}/\dot{M}_{w1}V_{w1}$ is the ratio between the wind momentum 
ratio of the two stars. Linder et al.\ (2009) suggest a distance-independent mass loss rate of 
$\sim2.1\times10^{-5}$ M$_\odot$ year$^{-1}$ for Cyg OB2 \#5 A. Its terminal velocity was 
suggested to be 1500 km s$^{-1}$ by Kennedy et al.\ (2010) based on the absorption of the 
P-Cygni He{\small I} 1.083 $\mu$m line observed by Linder et al.\ (2009). 

The value of $\eta$ will, of course, depend on the mass-loss rate and wind velocity of
Cyg OB2 \#5 B. In Table \ref{tab:pms} we summarize the parameters of massive stars with 
luminosity class V. The effective temperature, luminosity, radius, spectroscopic mass, and 
the flux of ionizing photons are taken from Vacca et al.\ (1996).  The mass loss rate was 
calculated following the prescription of Vink et al.\ (2001) for stars of solar metallicity, and the 
terminal velocity of the wind, $v_{\infty}$, was calculated assuming $v_\infty/v_{\rm esc}$ = 
2.6 (Vink et al.\ 2000), where $v_{\rm esc}$ is the escape velocity from the stellar surface. Finally, 
we calculated the expected flux density at 8.46 GHz assuming a fully ionized, pure hydrogen
wind at a distance of 1 kpc and following the formulation of Panagia \& Felli\ (1975). We
also used Equation 1 of Rodr\'\i guez \& Cant\'o\ (1983) to verify that the ionizing photon
rate of the stars was sufficient to fully ionize their respective winds.

\begin{table*}[!htbp]
\footnotesize
  \begin{center}
  \caption{Parameters of massive stars with luminosity class V. { The 4.8 GHz flux
  is estimated for a distance of 1 kpc. } Additionally, the value of $\eta$ 
  for each type of star is listed in the rightmost column, assuming that its wind would interact with 
  that of Cyg OB2 \#5 A.}
    \begin{tabular}{lccccccccc}\hline\hline
Spectral&T$_{\rm eff}$&$\log({\rm L/L}_\odot$)&$\log({\rm R/R}_\odot$)&$\log({\rm M/M}_\odot$)&$\log(N_i)$&$\log(\dot{M})$&v$_\infty$&S$_{8.46GHz}$&$\eta$\\
Type    &    (K)      &                       &                       &                       &           &(M$_\odot$ year$^{-1}$&(km s$^{-1}$)&(mJy)&     \\
\hline
  O3    &  51230  &  6.035  &   1.12  &        1.71   &   49.87 &  -4.90   &  3200   &   1.335  & 1.270 \\
  O4    &  48670  &  5.882  &   1.09  &        1.65   &   49.70 &  -5.13   &  3100   &   0.687  & 0.730 \\
  O4.5  &  47400  &  5.805  &   1.07  &        1.61   &   49.61 &  -5.24   &  3000   &   0.512  & 0.540 \\
  O5    &  46120  &  5.727  &   1.06  &        1.58   &   49.53 &  -5.36   &  2900   &   0.371  & 0.411 \\
  O5.5  &  44840  &  5.647  &   1.04  &        1.55   &   49.43 &  -5.49   &  2900   &   0.249  & 0.298 \\
  O6    &  43560  &  5.567  &   1.03  &        1.52   &   49.34 &  -5.63   &  2800   &   0.169  & 0.210 \\
  O6.5  &  42280  &  5.486  &   1.01  &        1.49   &   49.23 &  -5.77   &  2800   &   0.110  & 0.152 \\
  O7    &  41010  &  5.404  &   1.00  &        1.46   &   49.12 &  -5.92   &  2700   &   0.073  & 0.102 \\
  O7.5  &  39730  &  5.320  &   0.98  &        1.43   &   49.00 &  -6.08   &  2700   &   0.044  & 0.070 \\
  O8    &  38450  &  5.235  &   0.97  &        1.40   &   48.87 &  -6.24   &  2600   &   0.029  & 0.048 \\
  O8.5  &  37170  &  5.149  &   0.95  &        1.37   &   48.72 &  -6.41   &  2600   &   0.017  & 0.032 \\
  O9    &  35900  &  5.061  &   0.94  &        1.35   &   48.56 &  -6.61   &  2600   &   0.009  & 0.020 \\
  O9.5  &  34620  &  4.972  &   0.93  &        1.32   &   48.38 &  -6.79   &  2500   &   0.006  & 0.013 \\
  B0    &  33340  &  4.881  &   0.92  &        1.29   &   48.16 &  -6.99   &  2500   &   0.003  & 0.008 \\
  B0.5  &  32060  &  4.789  &   0.90  &        1.26   &   47.90 &  -7.21   &  2400   &   0.002  & 0.004 \\
\hline\hline
    \label{tab:pms}
    \end{tabular}
  \end{center}
\end{table*}

Cyg OB2 \#5 B must drive a wind strong enough to produce the WCR(A-B) when it interacts with 
the winds of the eclipsing binary Cyg OB2 \#5 A. As discussed by Kennedy et al.\ (2010) and
Ortiz--Le\'on et al.\ (2011), this suggests that Cyg OB2 \#5 B is of B0.5 or earlier spectral
type. The value of $\eta$ for a B0.5 star (Table  \ref{tab:pms}) yields an upper limit on the
distance to Cyg OB2 \#5 of 1.44 kpc. This result is, again,  in agreement with our measured 
trigonometric parallax, and with distances proposed by Hanson (2003), Rygl et al.\ (2012) 
and Zhang et al.\ (2012). 

Running the same argument in reverse, we can place an upper limit on the value of $\eta$ 
considering the highest possible mass for the system (90 M$_\odot$) and shortest possible
distance (1.26 kpc). This leads to $\eta$ = 0.047, corresponding to a spectral type O8 for 
Cyg OB2 \#5 B. This explains {\it a posteriori} why Cyg OB2 \#5 B contributed negligibly to the
thermal radio emission of the system: the total free-free emission from an O8 star at the 
distance of Cyg OB2 \#5 being only about 30 $\mu$Jy.

It is important to keep in mind that our results depend strongly on the orbital parameters reported 
by Kennedy et al.\ (2010). Fortunately, { once a model is selected, } their largest errors
are in the inclination of the orbit, while our calculations depend most strongly on the argument
of periastron and the eccentricity. 

\section{Conclusions}
In this paper, we presented an analysis of a series of multi--epoch VLBA observations of the 
compact wind collision region in the Cygnus OB2 \#5 quadruple system. The brightness 
temperatures for the most compact emission is $\sim10^7$ K, clearly suggestive of a non-thermal
process. The total measured flux  is consistent with a constant value of $\sim3$ mJy for the first 
seven epochs, but falls below the detection limit for the latest observations. This is consistent with
the interpretation proposed by Kennedy et al.\ (2010) that the shocked region plunges into the
photo-ionized region surrounding the primary in the system. In addition, in one of the early epochs,
the flux drops significantly. We suggest that for this epoch, a dense clump of the inhomogeneous 
wind intersect the sight of view, and the free--free emission partially hides the non-thermal emission 
from the wind collision region. 

The measured trigonometric parallax for the system corresponds to a distance of 1.65$_{-0.44}^{+0.96}$ kpc. 
In addition, the distance to the system was estimated using two indirect methods, both yielding distances
of order 1.3--1.4 kpc. These results are in agreement with recently measured trigonometrical parallaxes to
masers related with objects in the Cygnus X star forming complex, and discard the small value of 950 pc 
suggested by Linder et al.\ (2009).

A new set of VLBA observations when the WCR(A-B) comes back to its ''on'' state, combined with the positions 
presented in this work, could be used for a better determination of the trigonometrical parallax of the Cygnus 
OB2 \#5 quadruple system. We note that in agreement with Kennedy et al.\ (2010) the next ``on'' 
state will be on 2014 summer.

\acknowledgments
S.A.D., L.F.R., L.L., G.O.L. and A.T.A.\  acknowledge the financial support of 
DGAPA,  UNAM and CONACyT,  M\'exico. L.L.\ is indebted to the Alexander von Humboldt 
Stiftung for financial support. The National Radio Astronomy Observatory
is a facility of the National Science Foundation operated under cooperative 
agreement by Associated Universities, Inc.

\clearpage

\end{document}